# Variable-temperature scanning optical and force microscope

P. S. Fodor, H. Zhu, N. G. Patil, and J. Jevy

Department of Physics and Astronomy, University of Pittsburgh, Pittsburgh, PA 15260

**Abstract:**

The implementation of a scanning microscope capable of working in confocal, atomic force and apertureless near field configurations is presented. The microscope is designed to operate in the temperature range 4 – 300 K, using conventional helium flow cryostats. In AFM mode, the distance between the sample and an etched tungsten tip is controlled by a self-sensing piezoelectric tuning fork. The vertical position of both the AFM head and microscope objective can be accurately controlled using piezoelectric coarse approach motors. The scanning is performed using a compact XYZ stage, while the AFM and optical head are kept fixed, allowing scanning probe and optical measurements to be acquired simultaneously and in concert. The free optical axis of the microscope enables both reflection and transmission experiments to be performed.

I. INTRODUCTION

The invention of the scanning tunneling microscope (STM)[1] marked the beginning of a revolution in surface characterization and manipulation techniques, including atomic force microscopy (AFM), magnetic/electric force microscopy (MFM/EFM), scanning capacitance microscopy, and many others. The development of near field scanning optical microscopy (NSOM)[2] has revitalized the field of optical microscopy by enabling the diffraction limit to be surpassed. Increased spatial resolution is particularly important in studying nanoscale objects such as quantum wires, quantum dots and single molecules, where the properties of single structures are of interest. With NSOM, the limits imposed by the far-field excitation or detection are circumvented using small apertures such as the specially prepared fiber ends.[2] Higher spatial resolution has been achieved with so-called apertureless NSOM (ANSOM),[3] in which light is scattered from a scanning probe tip and detected in the far-field.

Here we report the development of a combined atomic force microscope (AFM) and high resolution optical microscope, designed to independently acquire force and optical measurements over a wide temperature range and at high magnetic fields. The system is both compact and modular, capable of operating in a variety of configurations, including ANSOM which gives the highest resolution of any optical technique. The working principle and design of the microscope will be described, along with representative data illustrating its capabilities.

## II. SYSTEM DESIGN

A perspective view of the microscope is depicted in Figure 1. The body of the microscope is cylindrical with a height of 14.50 cm and a diameter of 3.55 cm, and is designed to fit in a conventional helium flow cryostat. In particular, two cryostats are currently employed: a non-magnetic Oxford optistat bath cryostat and a Janis magneto-optical cryostat with maximum longitudinal and transversal fields of up to 2 Tesla and 8 Tesla, respectively. The sample chamber height difference between the cryostats is compensated using an extender tube which serves as both an electrical and mechanical interface. For both cryostats the temperature in the sample chamber is monitored using thin film resistance temperature sensors with an accuracy better then 10 mK at T < 10 K.

The microscope body is machined from a single piece of machinable glass ceramic Macor (Accuratus Ceramic Corp.). Besides its low weight and the capability to be machined easily with common metal tools, Macor has high mechanical and dielectric strength, allowing electrical wiring to be performed without supplementary isolation. More importantly, Macor has a low thermal expansion coefficient ($\alpha=9.3 \times 10^{-6}/°C$), which is comparable over the working temperature range to the other construction materials: titanium ($\alpha=8.6 \times 10^{-6}/°C$), stycast 1266 epoxy (Emerson & Cuning), and the piezoelectric ceramics ($\alpha=7.5 \times 10^{-6}/°C$) used for motion control.

The main parts of the microscope are an XYZ stage on which the sample is mounted, an AFM head operating above the sample and a high numerical aperture objective (NA=0.8, Partec) operating from below. Placing the optical objective in the cryostat is an important step in improving the optical spatial resolution. The vertical motion of the AFM head and objective is independently controlled using two inertial piezoelectric coarse approach motors. During scanning, the XY position of the AFM tip and optical objective are fixed, while the sample is scanned in-plane, allowing force and confocal optical images to be obtained simultaneously and independently. The optical resolution can be further increased using the apertureless near field microscopy (ANSOM) configuration, in which the optical image is reconstructed from the light scattered by the AFM tip.[3]

### A. AFM Module and Feedback Assembly

The AFM module of the microscope uses a quartz tuning fork sensor for feedback. This choice is motivated by two factors: (1) a desire to achieve a compact design compatible with multiple cryostats, while keeping the axis of the microscope open for both transmission and reflection experiments; and (2) the need to avoid artifacts from the scattered laser used for monitoring the cantilever motion.

The tuning forks used have a resonant frequency $f_0$ = 32 kHz and quality factors $Q \sim 10^4$ in air. The $Q$ degrades when the tip is attached to one of the prongs. However, quality factors for tip-equipped forks of $Q$>2000 are readily obtained in air at room temperature, with further increases as the microscope is cooled. Tips are produced by attaching a 12 μm tungsten wire to one of the tuning fork prongs. Silver epoxy is used so that bias voltages can be applied to the tip *via* the tuning fork electrode. The fork is then mounted on an electrochemical etching station, which uses a KOH electrolyte held in an iridium loop. The electrolysis reaction breaks away part of the tungsten wire leaving a tip with a radius smaller then 10 nm.[4] Since the tungsten tip is glued perpendicular to the tuning fork prong the sensor is operated in the tapping mode configuration,[5,6] rather than the shear mode originally employed with fiber tips attached along the length of the fork.[7-9] In tapping mode, the tip experiences higher force gradients which translate into increased resolution. Also, in the apertureless near field configuration, oscillation along the *z* direction (toward the sample) is the preferred mode for ANSOM.

In the tuning fork based AFM, the feedback loop can be implemented by sensing changes in the tuning fork resonance frequency and oscillation amplitude and phase with respect to the driving signal. Keeping one of these parameters constant allows a feedback loop to control the position of the tip relative to the sample.[6,7] In our set-up, the tuning fork is driven mechanically at resonance using a small piezoelectric mounted under the tuning fork. The deflection of the quartz tuning fork produces a piezoelectric response in the quartz. The phase of the response is maintained by changing the driving frequency with a phase-locked loop (EasyPLL, Nanosurf AG). The frequency shift required to keep the phase locked provides the feedback signal. The phase-locked loop allows the bandwidth of the system to be increased,[10-12] as compared to the constant frequency mode[7,13] where the high $Q$ translates into a slow response in the feedback loop.

At cryogenic temperatures, the signal-to-noise ratio is significantly reduced, due to the large capacitive load of the cables compared with the very small capacitance of the quartz tuning fork. For the tuning forks employed, the power loss at resonance is typically smaller than 10 nW. For optimal performance, a transimpedance amplifier is placed close to the sample. A single GaAs low noise field effect transistor (FET) (EPB018A5-70, Excelics Semiconductors) is employed, which can be operated over the entire temperature range without adjustment (Figure 2). This simple design allows bias voltages to be applied between the tip and the sample, which was not possible with a previously employed scheme using a symmetric pair of field-effect transistors.[14]

The frequency output from the phase locked loop is fed into a commercial scanning probe controller (PScan2™ controller, Pacific Scanning Corp.) which closes the feedback loop and acquires images. This tip-sample distance is regulated using a piezostack mounted above the AFM head. The stack is fabricated out of 14 washer-shaped piezo elements (Staveley NDT Technologies) poled along the stack axis. The 14 elements are divided into two groups containing 4 and 10 elements, respectively, so that three different ranges are accessible depending on the number of chips used. At higher temperatures, typically only 4 elements are required, while at the lowest operating temperatures all of the elements are used.

**B. XYZ Stage**

XYZ stage design was motivated by a need for compactness, and the ability to allow both AFM and optical access. The design is based on a planar one employing four quadrant piezotubes (PZT–5A, Diameter = 3.18 mm, Length = 15.24 mm, Boston Piezo Optics Inc.) previously proposed by Lieberman *et. al.*[15] The arrangement of the four piezotubes is depicted in Figure 3(a). One end of each tube is fixed rigidly in a Macor frame while the other is connected to an L shaped titanium plate holding a square titanium support plate. When opposite polarity voltages are applied to one pair of opposite electrodes, while the inner electrode is grounded, the piezotube will bend in the direction of the applied field. Thus, motion of the sample plate in the X or Y direction can be induced by applying voltages of equal magnitude to two opposite piezotubes (Figure 3(b)). Simultaneously bending all four tubes in the upward or downward direction leads to motion in the vertical direction. Balanced high voltage amplifiers are used to produce the control voltages for the quadrant piezotubes. The voltage signals controlling the XYZ motion are connected to the stage through 20 pins placed on the two sides of the Macor frame holding the piezotube assembly (Figure 3(a)). The pin assembly, together with two microwave connectors for RF sample biasing, provide a secure mechanical connection to the main frame of the microscope. The sample is mounted on a sapphire plate resting on top of the titanium support plate on three polished sapphire half-spheres. The half-spheres self-orient to match the planar surface of the sapphire sample plate, and provide a friction surface that allows slip-stick coarse motion in the XY plane.[16]

**C. Coarse Approach Motors**

The vertical position of the AFM head and the optical objective can be independently set using two similar coarse approach motors. The motor is based on a six piezostacks design holding a Macor inner shaft shaped like a triangular prism,[17] as shown in Figure 4(a). Each stack is made out of four rectangular shear plates, 10 mm x 8 mm x 1 mm (PZT 5A, Staveley NDT Technologies). Highly polished sapphire plates (0.25 mm thick) are glued on the stack and shaft surfaces to provide an even and flat mechanical contact. While two pairs of piezostacks are fixed in the microscope case, the third pair is tightened against the shaft using a titanium sheet wrapped around the motor. Changing the tension force on the titanium sheet through six screws allows the contact pressure to be adjusted. The force is applied at one point through a sapphire ball to allow even distribution. In order to insure that the stacks and shaft are flat and parallel to one another, the piezo stacks are epoxied to the microscope with the shaft in place allowing the entire motor to self-align.

When voltages are applied to the piezostacks they shear in the vertical direction. In order to translate the inner shaft, a voltage sequence is applied to the six stacks as shown in Figure 4(b). The voltage on each stack is ramped rapidly to a maximum value, set programmatically using a multi-channel 16-bit digital-to-analog PCI card. The voltage is then ramped down at a much slower rate. This allows a stick-slip motion of both the optical objective as well as the optical head to be performed. The translation motion is limited only by the physical length of the shafts.

## III. MICROSCOPE PERFORMANCE

The performance of the microscope was thoroughly tested over the entire temperature range, from 4 to 300 K. A primary concern was that the inertial coarse approach motors perform well over the entire temperature range (4 K – 300 K) and against gravity, without having to adjust the contact pressure. These conditions are met with the present design. The step amplitude versus the maximum applied voltage is presented in Figure 5 for two different temperatures. As expected, the step size at the same applied voltage decreases by a factor of three as the temperature is decreased to its lowest values, corresponding well to the decrease in the response of the piezo shear stacks. Reproducible steps as small as 10 nm are obtained both at room temperature, as well as at 4 K, both along and against the direction of gravity. The step size is consistently about two times smaller in the upwards direction, which is expected due to the downward force of gravity. Compensation schemes, such as using well-matched counterweights[18] are not required, and would also add to the design complexity.

Silicon calibration gratings are used to calibrate the motion of the XYZ stage and test its maximum range (Figure 6). At room temperature, a range of 6 μm is achieved in both X and Y directions. This value agrees well with the estimated value of about 5.8 μm for the quadrant piezotubes used at a maximum applied voltage of 200 V. The scanning range decreases monotonically as the temperature is lowered, in accordance with the temperature dependence of $d_{31}$ for PZT-5A. The limitations of the reduced scanning range at low temperatures, about 1 μm x 1 μm at 5 K, can be overcome using stick-slip coarse motion.[16] From the AFM images it can be seen that the pattern of the Si grating is well preserved. The width of the grooves is almost equal to that of the steps, showing that the tungsten tips used are sufficiently sharp to avoid artifacts arising from the convolution of the step with the sample surface structure.

Figure 7 shows independently obtained topography and confocal images of a two-dimensional quartz grating with hole dimensions 500 nm x 500 nm x 200 nm, spaced 1 μm apart. The confocal setup employs a beam expander with a pinhole to reject the scattered light coming from outside the focus of the in-situ high numerical aperture objective. Also, a balanced lock-in detection is used to reduce the noise background (Figure 8). During scanning, the z positions of the XYZ stage and the optical objective do not change, maintaining the grating always in focus. The optical resolution achieved is limited by diffraction, and for the laser light used, $\lambda = 638$ nm, is equal to about 500 nm. A discussion of the use of ANSOM with this microscope will be presented in a separate publication.

Atomic force microscopy images at low temperatures are obtained for a BaTiO$_3$ thin film grown epitaxially on a Si substrate, with a relaxed (BaSr)TiO$_3$ buffer layer. The topography image, at 10 K, shows well oriented elongated grains with a mean width of ~200 nm (Figure 9(a)). The local piezoelectric response of the sample is studied by applying DC bias voltages between the tip and the sample, while the tip is maintained in a constant position in the proximity of the sample. The shift in the resonant frequency of the tuning fork is interpreted as a measurement of the local polarization, similar to contrast in magnetic force microscopy. The hysteretic behavior, Figure 9(b), with saturation for large bias voltages corresponds to switching of the polarization along the c-axis, *i.e.*, normal to the film plane.

In summary, we have presented the design and implementation of a variable-temperature scanning probe/optical microscope capable of operation over a wide temperature range. We have demonstrated the capability of the microscope to simultaneously and independently perform force and optical measurements from room temperature down to 4 K. The ability to spatially correlate the structural and optical properties is particularly important in understanding and manipulating semiconductor heterostructures, such as quantum dots, whose properties are dominated by the quantum confining effects.

**Acknowledgments**

The authors are extremely thankful to Bob Giles for his help in machining the parts necessary for the microscope, and to Darrell G. Schlom for providing the ferroelectric sample. This work was supported by NSF (9802784) and DARPA (DAAD19-01-1-0650).

**FIGURE CAPTIONS**

**Figure 1** Schematic drawing of the scanning microscope. (1) upper optical port, (2) lower optical port, (3) Macor case, (4) AFM head coarse approach assembly, (5) tuning fork sensor, (6) Ti sheet for contact pressure adjustment in the coarse approach motors, (7) sapphire ball, (8) coarse approach motor shaft, (9) optical objective, (10) XYZ stage, (11) microwave connectors, (12) piezostacks.

**Figure 2** Schematics of the in-situ single FET amplifier.

**Figure 3** a) Perspective view of the XYZ stage. (1) quadrant piezotubes, (2) titanium base, (3) sapphire sample plate, (4) Macor base, (5) L shaped titanium plates, (6) microwave connectors, (7) electrical pins, (8) macor frame. b) Opposite voltages applied to electrode pairs while the center electrode is grounded, bend the piezotube in the field direction.

**Figure 4** a) Side and bottom view of the objective coarse approach motor. (1) Macor case, (2) sapphire plates, (3) piezostacks, (4) shaft. b) Sequence of voltages applied to the six piezostacks.

**Figure 5** Amplitude of coarse approach step vs. maximum applied voltage at T = 220 K (■ and ●) and 5 K (◀ and ⬧), respectively. Positive voltages correspond to motion against gravity.

**Figure 6** Temperature performance of the XYZ stage.

**Figure 7** Images obtained using (a) AFM and (b) confocal scanning optical microscopy of a quartz calibration grating at room temperature.

**Figure 8** Confocal setup with a balanced photodetector and lock-in detection

**Figure 9**   a) AFM image of a strain BaTiO$_3$ thin film grown on Si, taken at 10 K. b) Piezoresponse signal obtained by cycling the voltage between the tip and the sample.

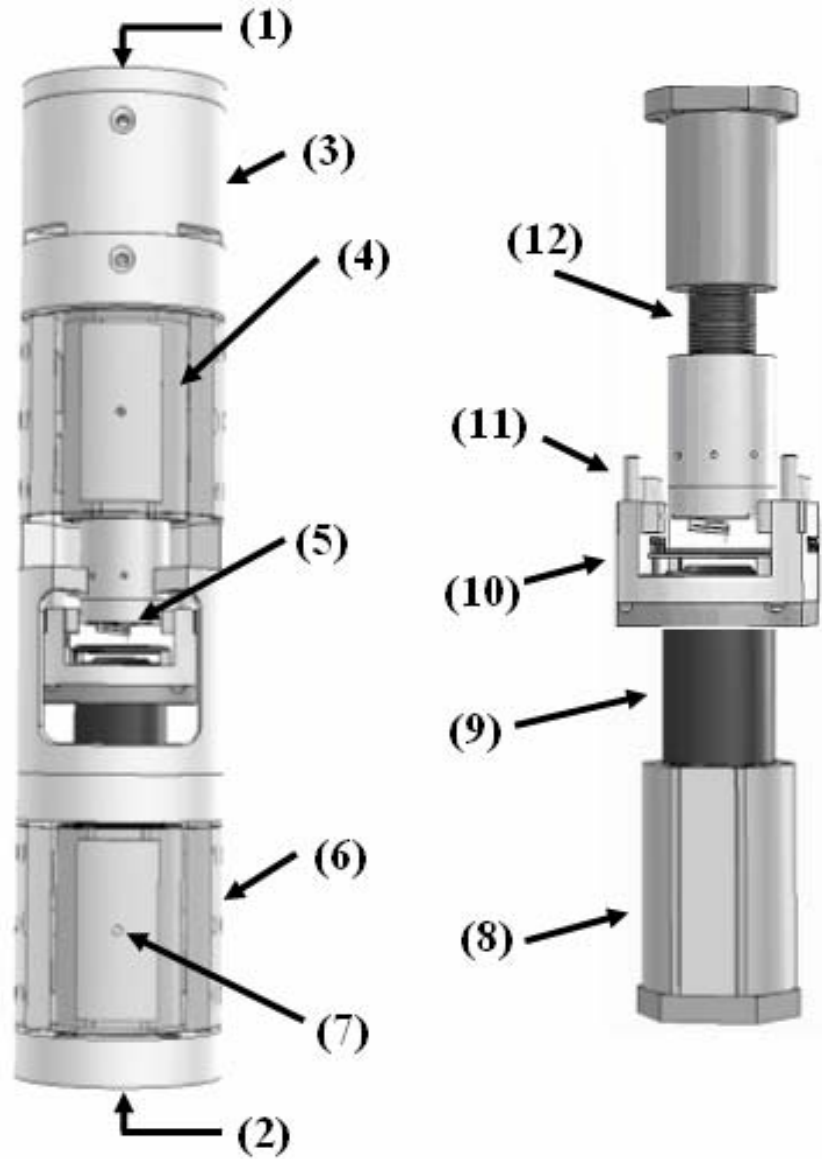

**Figure 1**  Schematic drawing of the scanning microscope. (1) upper optical port, (2) lower optical port, (3) Macor case, (4) AFM head coarse approach assembly, (5) tuning fork sensor, (6) Ti sheet for contact pressure adjustment in the coarse approach motors, (7) sapphire ball, (8) coarse approach motor shaft, (9) optical objective, (10) XYZ stage, (11) microwave connectors, (12) piezostacks.

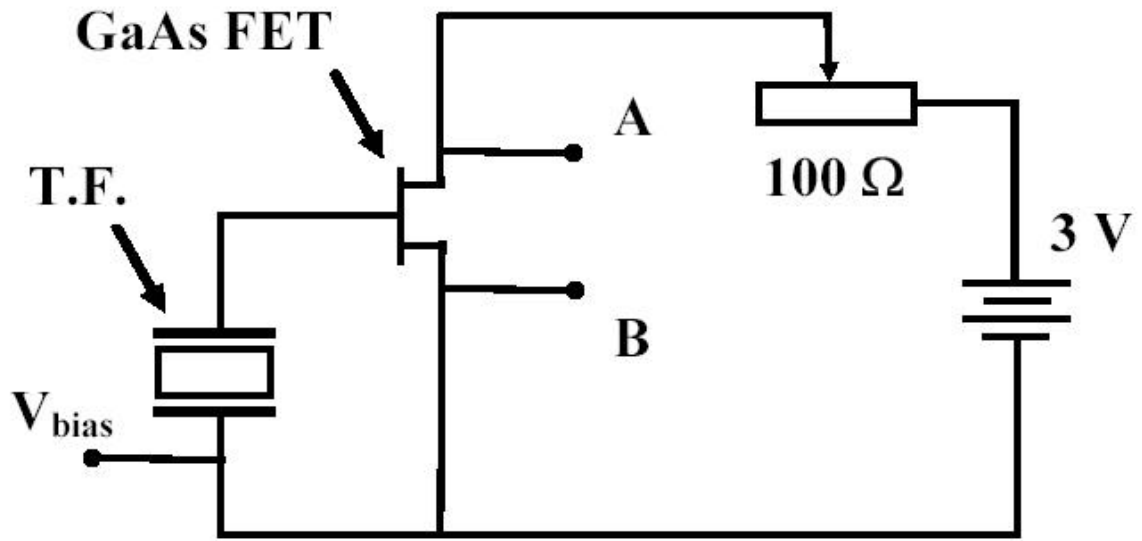

**Figure 2**     Schematics of the single FET amplifier.

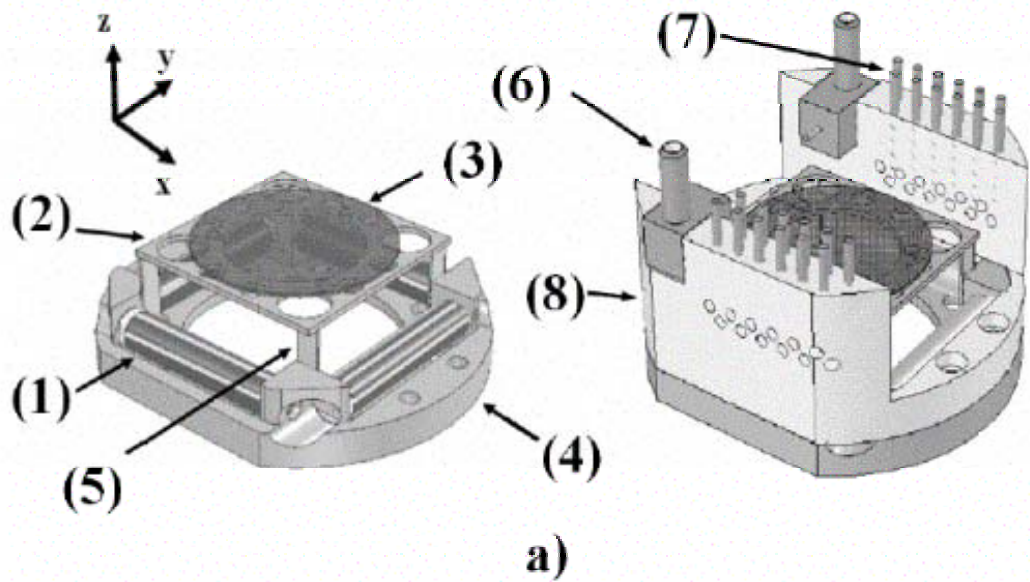

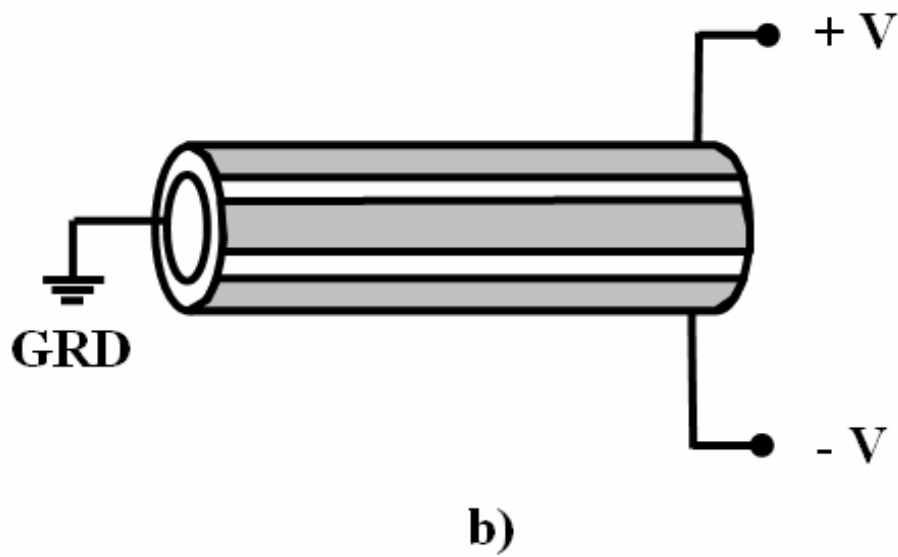

**Figure 3**  a) Perspective view of the XYZ stage. (1) quadrant piezotubes, (2) titanium base, (3) sapphire sample plate, (4) Macor base, (5) L shaped titanium plates, (6) microwave connectors, (7) electrical pins, (8) macor frame. b) Opposite voltages applied to electrode pairs while the center electrode is grounded, bend the piezotube in the field direction.

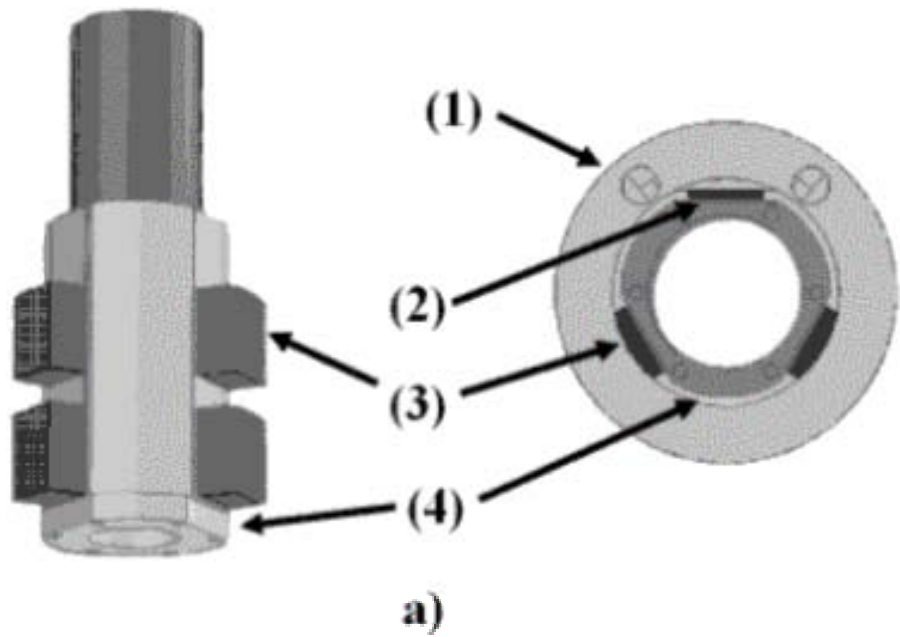

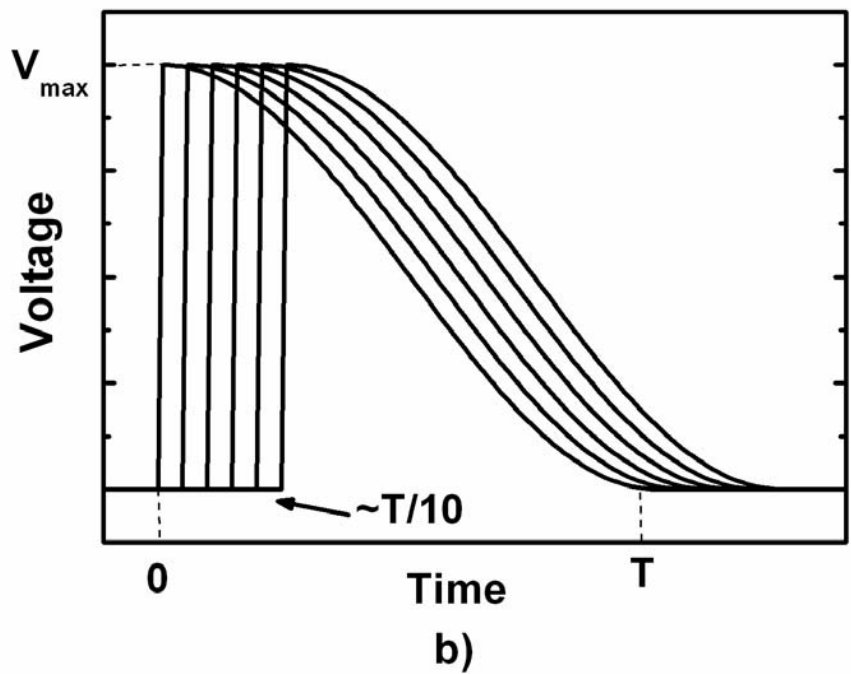

**Figure 4**  a) Side and bottom view of the objective coarse approach motor. (1) Macor case, (2) sapphire plates, (3) piezostacks, (4) shaft. b) Sequence of voltages applied to the six piezostacks.

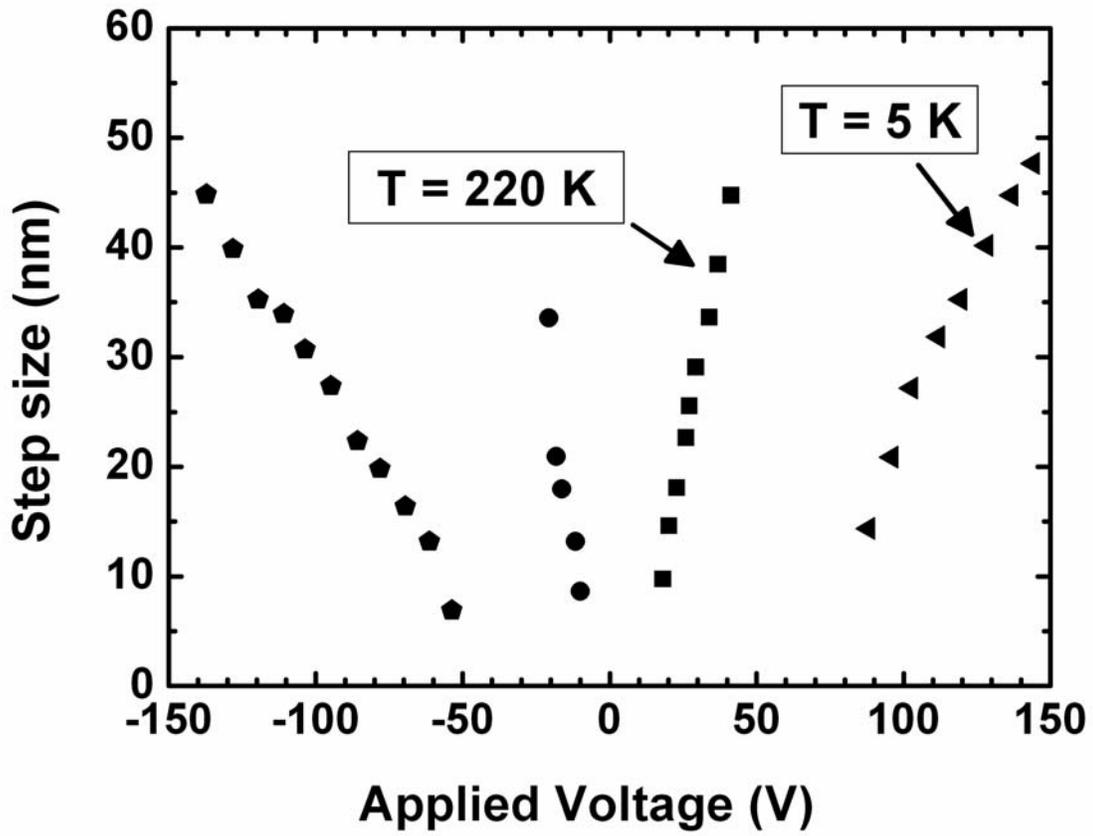

**Figure 5**  Amplitude of coarse approach step vs. maximum applied voltage at T = 220 K (■ and ●) and 5 K (◀ and ⬟), respectively. Positive voltages correspond to motion against gravity.

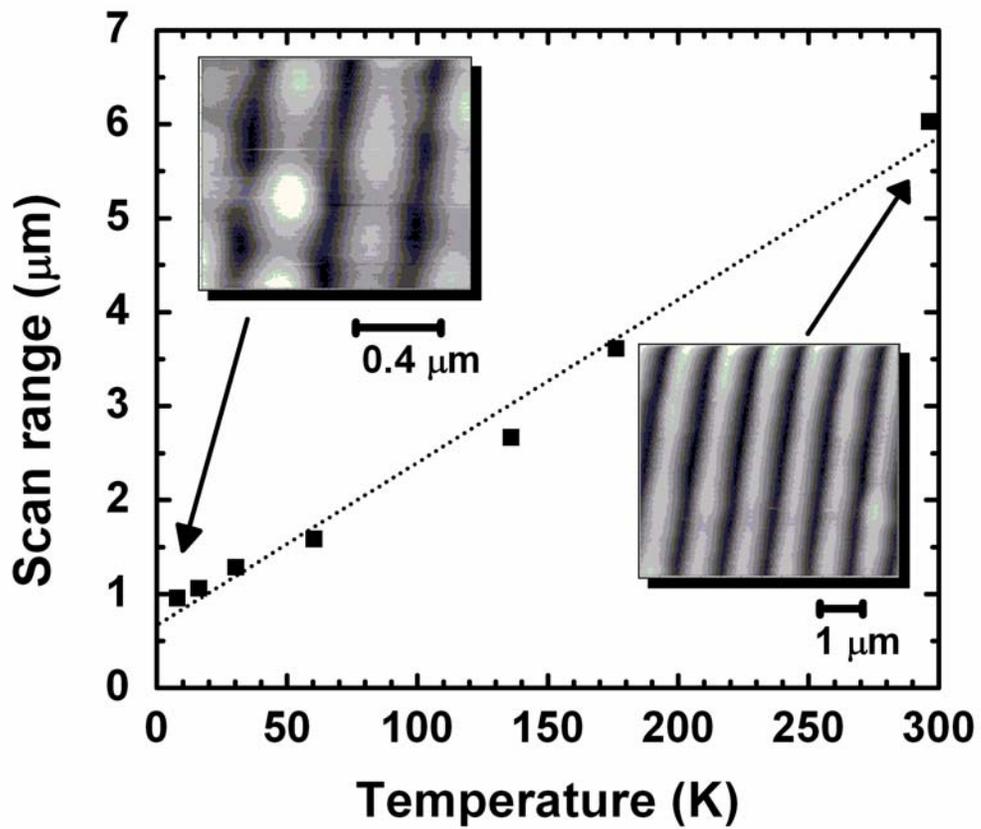

**Figure 6** Temperature performance of the XYZ stage.

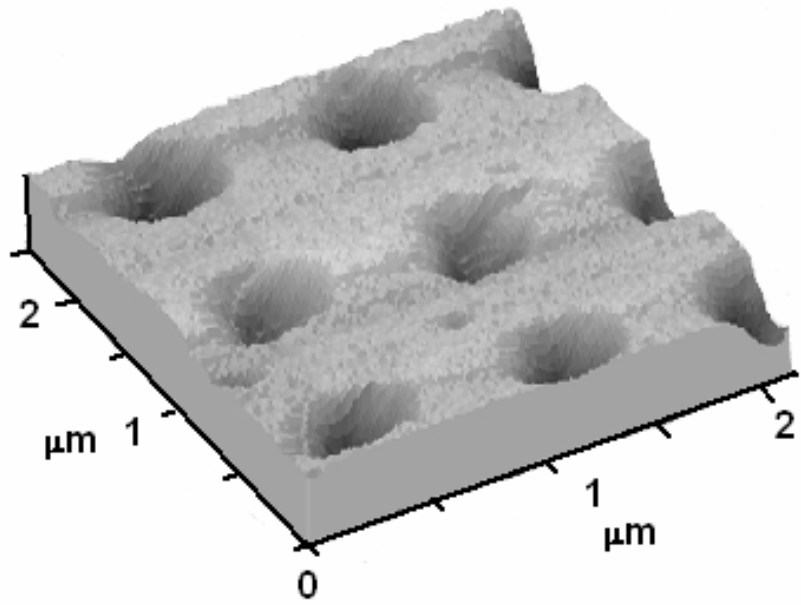

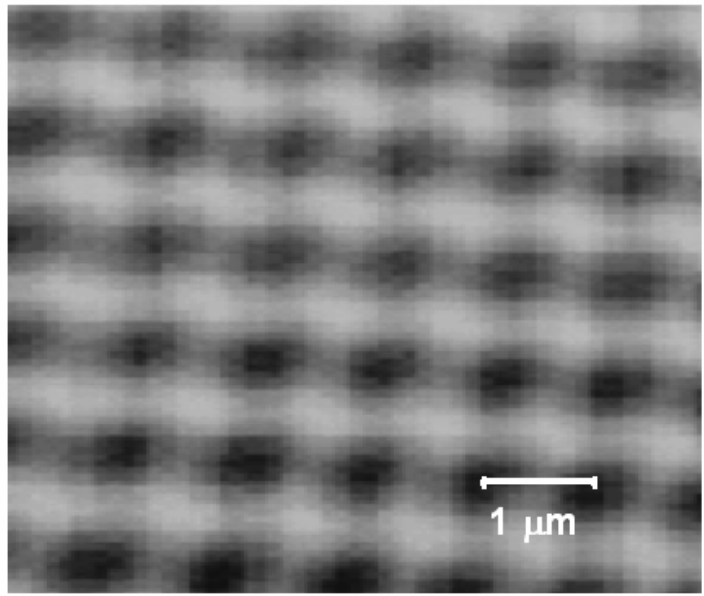

**Figure 7**  Images obtained using (a) AFM and (b) confocal scanning optical microscopy of a quartz calibration grating at room temperature.

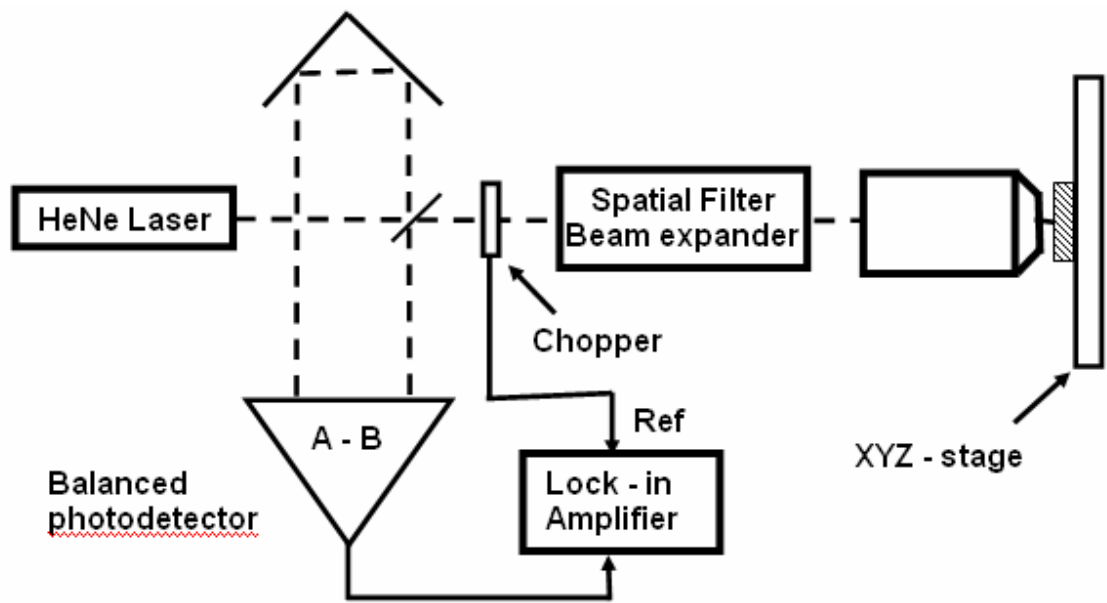

**Figure 8** Confocal setup with a balanced photodetector and lock-in detection

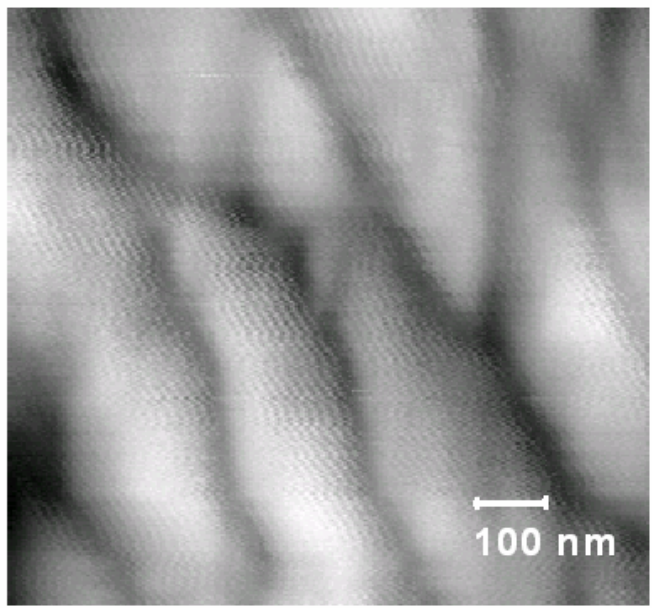

a)

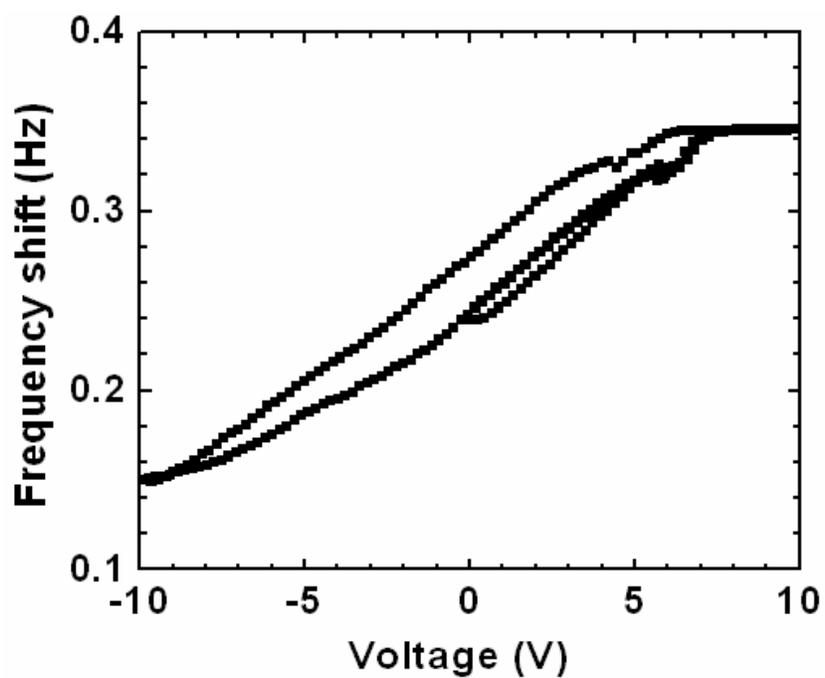

b)

**Figure 9**     a) AFM image of a strain BaTiO$_3$ thin film grown on Si, taken at 10 K. b) Piezoresponse signal obtained by cycling the voltage between the tip and the sample.